\documentclass[%
 reprint,
 amsmath,amssymb,
 aps,
]{revtex4-2}

\usepackage{graphicx}
\usepackage{dcolumn}
\usepackage{bm}

\usepackage{xr-hyper}
\usepackage[colorlinks]{hyperref}
\hypersetup{
colorlinks=true,
linkcolor=blue,
urlcolor=blue,
filecolor=blue,
citecolor=blue
}

\usepackage{empheq}
\usepackage{tabularx}
\usepackage{subfigure}
\usepackage{scalerel,amssymb}
\usepackage{textcomp, gensymb}

\DeclareMathAlphabet{\mathcal}{OMS}{cmsy}{m}{n}

\usepackage{xcolor}
\usepackage[utf8]{inputenc}
\usepackage[T1]{fontenc}
  
\usepackage[normalem]{ulem}
\usepackage{times}

\begin{document}

\preprint{APS/123-QED}

\title{Critical role of the motor density and distribution on polar active polymers}

\author{Surabhi Jaiswal}
\email{surabhi19@iiserb.ac.in}
\affiliation{Department of Physics, Indian Institute of Science Education and Research Bhopal, Bhopal 462066, India}

\author{Prithwiraj Maity}
\affiliation{Department of Physics, Indian Institute of Science Education and Research Bhopal, Bhopal 462066, India}

\author{Snigdha Thakur}
\email{sthakur@iiserb.ac.in}
\affiliation{Department of Physics, Indian Institute of Science Education and Research Bhopal, Bhopal 462066, India}
\author{Marisol Ripoll}
\email{m.ripoll@fz-juelich.de}
\affiliation{Theoretical Physics of Living Matter, Institute for Advanced Simulation, 
Forschungszentrum J\"{u}lich, 52425 J\"{u}lich, Germany}

\date{\today}

\begin{abstract}
Polar polymer activity is a fundamental mechanism behind a large number of cellular dynamical processes. The number and location of the active sites on the polymer backbone play a central role in their dynamics and conformational properties. 
Globular conformations for high motor densities change to stretched ones for the more realistic moderate or low density of motors, with a self-propelled polymer velocity non-monotonically related to the motor density. A small difference in the position of the first motor, or the motor distribution, can also dramatically modify the polymer typical conformations. 
\end{abstract}

\maketitle


A large number of crucial biological functions in the cell are governed by polymer activity.
Biological polymers become active because of the presence of molecular motors or enzymes, and synthetic polymers might come into contact with active baths or with external activation, such as magnetism or phoretic processes.  
The genome is organized into highly dynamic chromatin structures that regulate the access to progressive enzymes such as RNA polymerase to transcript encoded genes~\cite{lewin2008genes,zhang2015structural, peter1988transcription,  bai2006singlemolecule}.
Cellular filaments such as actin or microtubules move by the action of cytoskeletal molecular motors such as kinesin or myosin~\cite{mizuno2007nonequilibrium, humphrey2002active, ahmet2004kinesin}. In the process, such molecular machinery consumes chemical energy and exerts mechanical forces on the polymer~\cite{zhou2016motors,cao2024idealchromosome,needleman2017active, van2019role, struhl2013determinants, fudenberg2016loopextrusion,dogterom2019actin,lopez2014actin}.
The role of active forces on flexible polymers like chromatin in the context of genome organization, or in semiflexible filaments like actin, microtubule are currently being investigated using a variety of experimental techniques~\cite{andriy2023polymer, adrian2015chromatin,lieberman2009comprehensive,heermann2011physical,julio2009spatially,sachs1995randomwalk,rao20143dmap,wen2011polymer,anderson2024biopolymer}. 

Active polymers have been investigated also by a number of simulation and analytical mesoscopic approaches that incorporate motor activity in different flavors. Driving the polymer out of equilibrium by imposing different temperature at different monomeric positions~\cite{agarwal2020nonequilibrium,smrek2020active}, employing colored noise~\cite{ghosh2014semiflexible,dino2017rousechains,Brahmachari2024temporally}, or data driven procedures where effective equilibrium polymer models optimize the inter-monomer interactions to generate structures consistent with the experimental observations~\cite{quinn2018chromatin,chiang2019chromosome,guang2021hic,shin2022chromatinphase}. Other type of schemes consider the non-equilibrium aspects of how self-propulsion force is locally applied. This can be either without any correlation along the backbone (active Brownian polymers)~\cite{kaiser2015flexiblechainswell,mousavi2019active, martin2019active, gandikota2022effective,mousavi2021active} or parallel to the backbone (polar active polymers)~\cite{bianco2018activepol,foglino2019motors,bianco2021Activity-Induced,xiang2023polarpol,sandeep2023local,tejedor2024progressive,oller2025melt}. 
In particular, the latter forcing model is similar to the sliding of RNA polymerase on chromatin during transcription~\cite{debora2023rnasliding,bruce2002molecularbiology,zia2011modeling} or molecular motor walking on the microtubule~\cite{wen2011polymer,anderson2024biopolymer}. 
The polar activity on the polymer shows interesting properties that depends on its flexibility as well as its dimensionality~\cite{schaller2010polar,sciortino2023polarity,bianco2018activepol,pratyusha2018dynamically,vatin2024conformation,isele2015selfpropelled}. 

Most of the theoretical understanding on polar active polymers has until now focused on the case where all constituent monomers are active. With this constraint, it has been shown that the introduction of polar activity systematically reduces the extension of the polymers resulting even on globule-like structures~\cite{bianco2018activepol,surabhi2024bd,namita2022collapse}. However, for experimental bio-polymers, as well as synthetic polymers, activity is typically exerted just on some of the monomers. For example, chromatin transcriptional activity lies within a few segments only along the genome, the genes, and it turns out that approximately $20\%$ of the total genes are transcriptionally active at a given time, leading to only a fraction of the genome experiencing transcriptional activity carried out by RNA polymerase~\cite{lewin2008genes}. 
Cytoskeletal molecular motors like kinesin or myosin attach only to a few spots in each actin or microtubule. Similarly, in the case of synthetic polymers varying the polymer properties by the motor density offers a promising avenue for developing new applications.   
The role played by the density of the motors and their position is therefore a fundamentally important aspect to be investigated. 

\begin{figure}[ht!]
     \includegraphics[width=\linewidth]{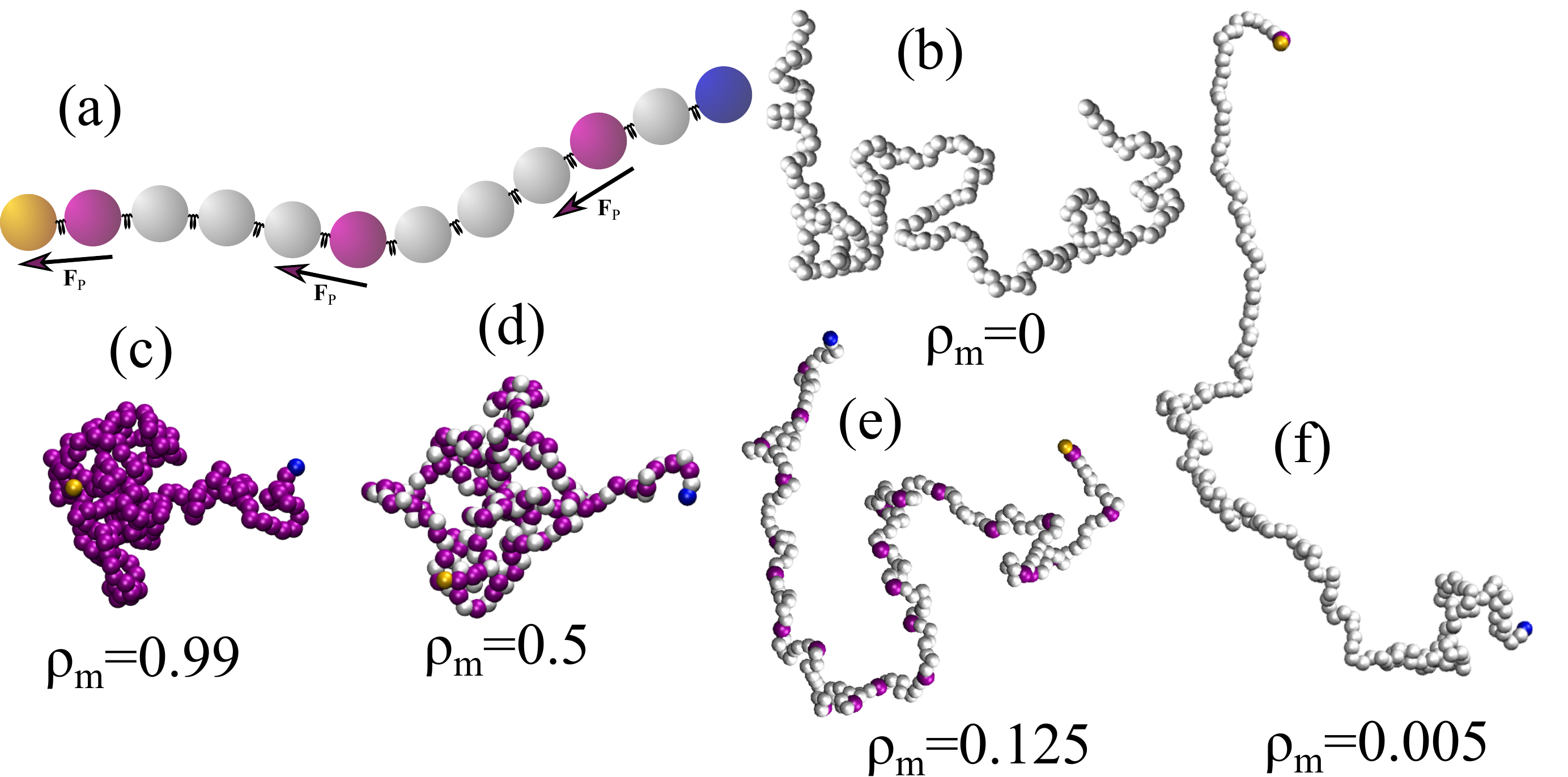}
     \caption{Polar polymer structure and typical configurations. (a) Polymer sketch with head bead in yellow, motor beads in purple, tail bead in blue, and linker beads in white. The arrows indicate the direction of the polar active force. Snapshot for the passive polymer is depicted for reference in~(b), while snapshots from~(c) to~(f) illustrate globule to stretch conformations with decreasing motor densities ($\rho_m$) for the chain length  $N=200$ at $\textrm{Pe}=17$. See Movie1 and Movie2 in SM.}
     \label{fig:polsketch}
 \end{figure}
Linear flexible polymers with polar activity are investigated here with a simulation approach provided by the overdamped Langevin equation where hydrodynamics are not considered~\cite{surabhi2024bd, roca2022self, pathria2016statistical}.
Each polymer is composed by $N$ beads with diameter $\sigma$, and consecutive beads connected with harmonic springs and excluded volume interaction among the non-neighboring beads, introduced through the repulsive WCA potential (see supplemental material for details). The monomers selected as motors have a force ${\bf F}_P=-f_c{\hat {\bf e}}_{j} $ pointing toward the previous bead, as depicted in Fig.~\ref{fig:polsketch}(a). The unit vector ${\hat {\bf e}}_{j} =({\bf r}_j-{\bf r}_{j-1})/|{\bf r}_j-{\bf r}_{j-1}|$ with ${\bf r}_j$ the position of motor $j$.  The number of motors $N_m$ can therefore vary between $0$ and $N-1$, such that the density of the motors is $\rho_m=N_m/N$, with $\rho_m=0$ corresponding to the passive polymer, see Fig.~\ref{fig:polsketch}(b) and $\rho_m=0.99$ to the case where all beads but the first are active (see Fig.~\ref{fig:polsketch}(c)). 
First, the motors are uniformly distributed along the polymer, and then we also consider the non-uniform distribution case.  A random initial configuration is considered and the simulation is run until steady state is reached. 
Activity can be characterized in terms of the monomeric P\'eclet number, $\text{Pe}={f_cb}/{k_\textrm{B}T}$, which refers to the ratio of locally applied active force and thermal fluctuations. Here, $k_\textrm{B}$ is the Boltzmann constant, $T$ is the temperature of the system. 

 \begin{figure}[ht!]
     \includegraphics[width=\linewidth]{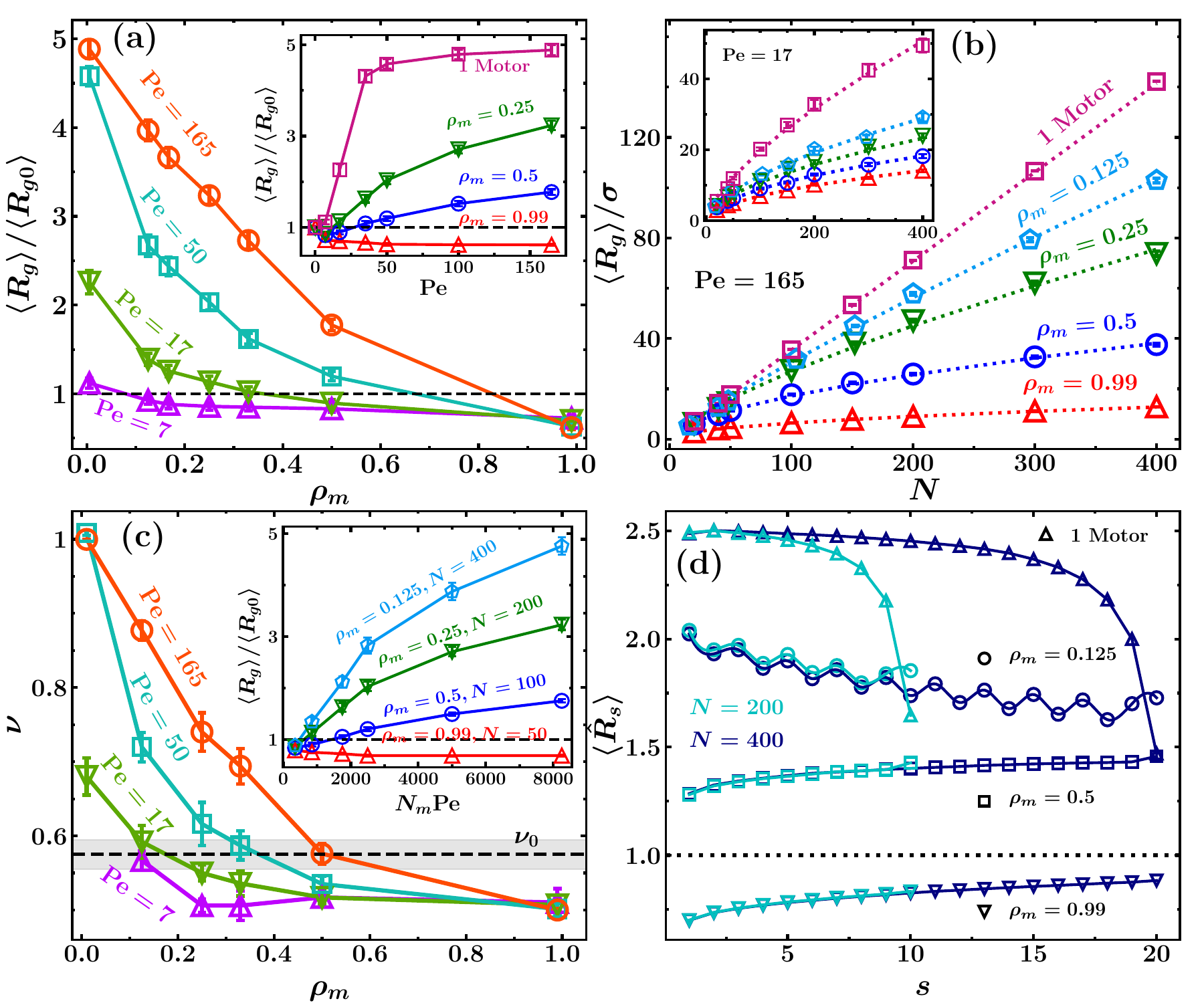}
     \caption{Conformational stretching and collapsing polymer properties.  
     (a)~Averaged radius of gyration normalized by the equilibrium value as a function of the motor density in the main plot and of the P{\'e}clet number in the inset for $N=200$. 
              (b)~Polymer size dependence of $\langle R_g\rangle$ for $\textrm{Pe}=165$ in the main plot and $\textrm{Pe}=17$ in the inset. Dotted lines are a fit to  $\langle R_g\rangle \sim N^{\nu}$  to determine $\nu$. 
              (c)~Flory exponent~($\nu$) as a function of motor density $\rho_m$ for various Pe values.
              The inset shows $\langle R_g\rangle$ as a function of the total applied force $N_m$Pe in four comparable cases. 
               (d)~Mean normalized segment size along the chain contour, with $s$ the segment number and $s=1$ corresponding to the head for $\textrm{Pe}=50$, for various motor densities and two values of the polymer length. In (a), (c) and (d) solid lines are a guide to the eye and the dashed line is the reference to the equilibrium value. See Movie1 and Movie2 in SM.}
     \label{fig:rgasdpe}
 \end{figure}

In the limit of full activity, the polymers have been characterized to coil into structures more compact than equilibrium, while in the single head motor limit the structures get stretched~\cite{bianco2018activepol, tejedor2024progressive,natali2020headactive,xiang2023polarpol}, as shown also in Fig.~\ref{fig:polsketch}. 
The intermediate densities get compactified or stretched depending not only on $\rho_m$ but also on Pe, which occurs independently on the polymer length.  
In order to quantify the conformational changes, we calculate the radius of gyration $R^2_{g} \equiv \sum_{i=1}^{N}\left({\bf r}_{i}-{\bf r}_{cm}\right)^2/N$, where ${\bf r}_{cm}$ is the center of mass position of the polymer, and ${\bf r}_i$ the position of $i^{\textrm{th}}$ bead. 
Figure~\ref{fig:rgasdpe}(a) shows the averaged $R_{g}$ in steady-state for different motor densities and activities. The data is normalized with the value at equilibrium, $R_{g0}$, which makes a clear distinction between compactified and stretched configurations.
Small activity values show to compactify the polymers, opposite to larger activities for which the stretching increases when increasing activity and also when reducing the number of motors, with the only exception of the all beads-motor case, where compactification increases with activity.
The dependence of the radius of gyration with the polymer size is shown in Fig.~\ref{fig:rgasdpe}(b), where a clear power law dependence $\langle R_g\rangle \sim N^{\nu}$ can be identified for each value of the motor density and the P{\'e}clet, with values summarized in Fig.~\ref{fig:rgasdpe}(c). Note that the one motor case does not correspond to a well-defined density, such that we have approximated this limiting case with $\rho_m \simeq 0.01$. 
In equilibrium, the exponent $\nu$ is known as the Flory exponent and a theoretical prediction simply considering a balance of entropic elasticity and excluded volume interactions which for a three dimensional polymer results in $\nu_0=0.588$~\cite{doi1988theory}. 
We measure here $\nu_0=0.575\pm0.02$, which agrees within the error with the theoretical prediction. 
A perfectly compact polymer has $\nu_c=0.33$ by construction, and a perfectly stiff polymer $\nu_s=1$. This seems to indicate that cases with $\nu < \nu_0$ should correspond to compact states, while cases with  $\nu > \nu_0$ should relate to stretched states. 
Previous studies of polar polymers in the all motor case have verified this extent with compact structures with  $\nu < \nu_0$~\cite{bianco2018activepol,namita2022collapse}. 
Our results in Fig.~\ref{fig:rgasdpe}(c) show that this is a good estimation, but interestingly, it is not always exactly fulfilled. This means that there are a few cases (e.g. $\rho_m=0.25$, $\text{Pe}=17$) for which this is not exactly the case, and the polymers are getting a slightly stretched with respect to equilibrium at the same time that their effective size increase with increasing $N$ is smaller than equilibrium. 

Besides the monomeric P\'{e}clet number, the activity can be evaluated by the total force applied on a polymer, this is $N_m{\textrm {Pe}}$, which can also be related to the total energy consumption. For a given size of the polymer, this total force can be varied either by changing $\rho_m$ or by changing $\textrm{Pe}$. 
In principle, an overall effective stretching could have been directly related to this total force, but the inset in Fig.~\ref{fig:rgasdpe}(c) shows 
otherwise, and makes obvious that with the same energy consumption and the same number of motors the lines do not collapse into a single one, such that $N_m$Pe does not determine the overall stretching, and the stretching is larger the larger the chain. 
Similarly, the inset of Fig.~\ref{fig:rgasdpe}(a) already shows that given the same activity and the same polymer length, the stretching is larger the smaller the number of motors.  

Local conformations have also shown to be importantly affected by polar activity in the single polymer case~\cite{tejedor2024progressive}, as well as in melt conditions~\cite{oller2025melt}. Figure~\ref{fig:rgasdpe}(d) depicts $\langle \hat{R}_s \rangle$, the  end to end distance of polymer segments with $N_s=20$ monomers along the contour length, normalized with the segment size in equilibrium.  
For the all-motor case, $\rho_m=0.99$,  the head is clearly more compact than the tail, trend which is maintained for the case where every second bead is active,  $\rho_m=0.5$, although in this case the segments have stretched with respect to equilibrium, $\langle \hat{R}_s \rangle > 1$, in contrast to the all active case where they were the segments have got more compact, $\langle \hat{R}_s \rangle < 1$. 
A further decrease in motor density reverses this trend and the stretching monotonously decreases toward the tail, an effect that becomes more prominent the smaller the number of motors (the oscillations for $\rho_m=0.125$ are due to the different number of motors in consecutive segments).
The overall behavior of $\langle \hat{R}_s \rangle$ is independent of $N$, besides some deviation for the last segment due to the difference in tension. The main conclusions drawn here are qualitatively the same with different values of the applied activities, (see Fig.~S1 in SM). 
Therefore, polar activity breaks down the polymer self-similar scaling behavior in a very consistent manner.

\begin{figure}[ht!]
    \includegraphics[width=\linewidth]{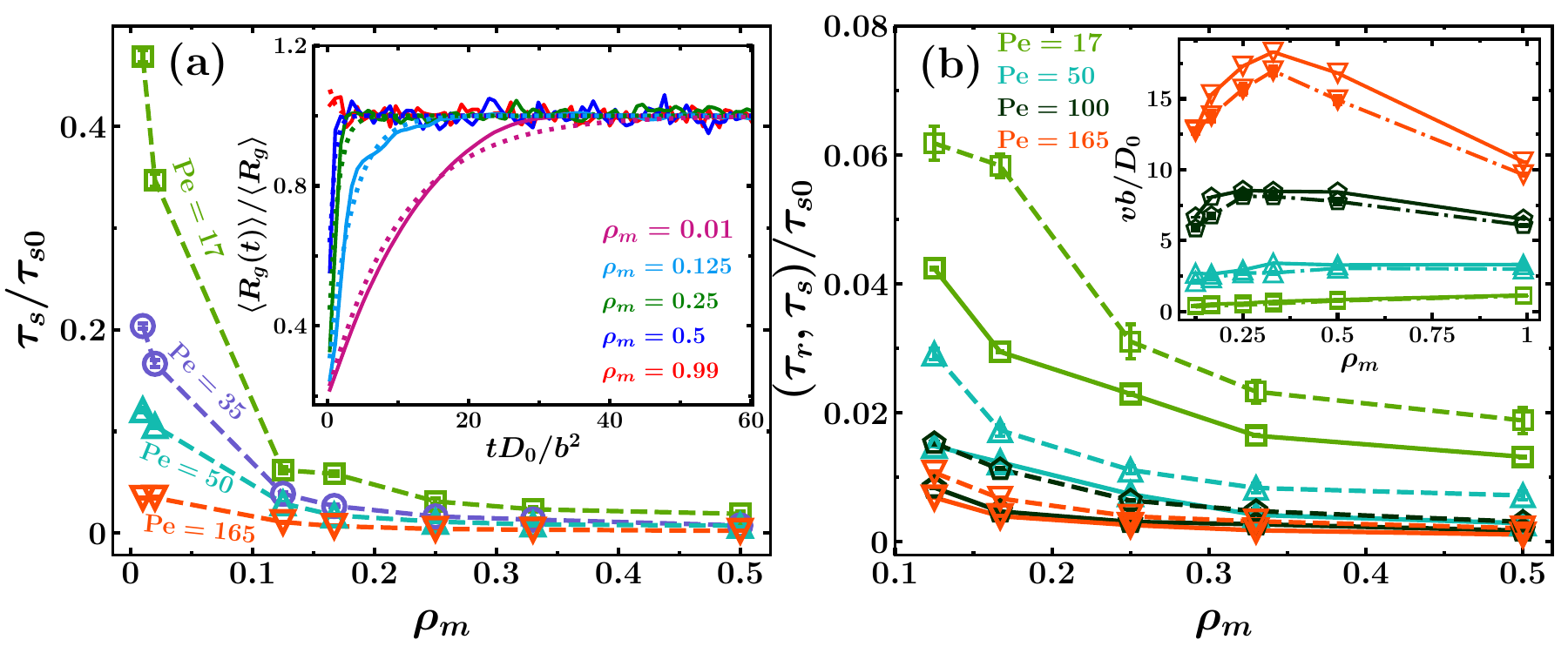}
     \caption{Dynamic polar polymer properties.
     (a)~Characteristic steady-configurational time $\tau_s$ as a function of motor density, for $N=100$, normalized by the equilibrium value $\tau_{s0}$. The inset displays time dependent average radius of gyration for various motor densities, where the dotted lines correspond to Eq.~(\ref{eq:radiusofgy}), with $\tau_s$ as fitting parameter, for $N=100$ and $\textrm{Pe}=165$. (b)~Characteristic rotational time $\tau_r$ as obtained from the MSD for $N=400$ (solid lines), compared with a zoom in of the data in (a) for  $\tau_s$ (dashed lines).  The inset corresponds to the polymer center of mass velocity as obtained from the MSD for $N=400$ (solid lines). Dashed-dotted lines in the inset correspond to the velocity estimation in terms of the effective force as, $ v b/D_0=\rho_m PeR_e/(Nb)$, where $R_e$ is the end-to-end polymer distance. Time and velocity are given in units of the monomers bond length $b$, and the equilibrium single monomer diffusion coefficient $D_0$ (see details in SM). See Movie1 and Movie2 in SM.
     }
     \label{fig:dynamics}
 \end{figure} 
 The dynamic behavior of the polar active polymer is also importantly affected by the motor density. We quantify this effect in two ways. 
First, the time evolution from an arbitrary initial state to the final steady-state under polar activity, allows us to determine the characteristic steady-configurational time $\tau_s$ with,
\begin{equation}
\label{eq:radiusofgy}
\frac{\langle R_g(t) \rangle}{\langle R_g\rangle}= 1 - a_1\exp[-(t/\tau_s)],  
\end{equation}
where  $a_1= 1-R_g(0) / \langle R_g\rangle$,
 is a constant determined by the initial configuration $R_g(0)$, such that $\tau_s$ is now the only fitting parameter~\cite{paul2022alignmentactivity}. The time evolution of the normalized radius of gyration is displayed in the inset of Fig.~\ref{fig:dynamics}(a) for different motor densities at $\textrm{Pe}=165$ for $N=100$. Higher density of motors imparts the higher activity to the polymer making the evolution faster and therefore reducing $\tau_s$.  Figure~\ref{fig:dynamics}(a) shows that $\tau_s$ decreases with increasing $\textrm{Pe}$ and $\rho_m$, this is with increasing local activity and motor density. 

 The second characterization of the polymer dynamics is performed through the center of mass mean square displacement (MSD) calculated in the steady-state for different densities and $\textrm{Pe}$ values. The MSD shows the typical behavior of an active Brownian particle (ABP)~\cite{andreas2016activecolloids,howse2007selfmotile,tejedor2024progressive} with a long time ballistic regime followed by a diffusive regime (see Fig.~S2 in SM).
 This enables the determination of the polymer self-propulsion velocity $v$ and the characteristic rotational time $\tau_r$, both displayed in Fig.~\ref{fig:dynamics}(b). 
The characteristic rotational time decreases with increasing Pe, which relates to the faster reorientation of the trajectories with increasing activity. Increasing the density of motors also decreases the local and global reorientation of the polymer which results in smaller values of $\tau_r$. 
For comparison, Fig.~\ref{fig:dynamics}(b) also includes a zoom in of the values for $\tau_s$, which show to be very similar. The differences can be related to variation of polymer length between the two data sets.
The fact that the two characteristic times $\tau_s$ and $\tau_r$ are essentially the same indicates that in the presence of polar activity there is a unique typical characteristic time which decreases with activity and motor density.  

The self-propulsion velocity $v$, in the inset of Fig.~\ref{fig:dynamics}(b), increases monotonically with $\textrm{Pe}$ for all densities, as expected. On the other hand, the dependence with motor density for a given $\textrm{Pe}$ is non-monotonic. 
To understand this effect, the effective force acting on the polymer center of mass, $F_e$, needs to be considered. 
The total applied force, $N_m$Pe, increases both with $\rho_m$ and Pe, but this force directly determines the polymer velocity only in the case of total stretching. For all polymer structures, the velocity is determined by $F_e$ and the total polymer friction, this is $v = F_e/(N\mu)$ with $\mu=k_\textrm{B}T/D_0$ the friction on a single monomer, and $F_e=\rho_m R_e$Pe, with $R_e$ the measured polymer end-to-end distance. The inset of Fig.~\ref{fig:dynamics}(b) shows that this approximation explains the velocity dependence very well.  For large values of the activity, maximum propulsion is therefore achieved at a particular intermediate motor density. Decreasing the value of the activity, the non-monotonicity also decreases, and for relatively small Pe values the velocity increases slowly and monotonically  with increasing motor density.

\begin{figure}[ht!]
\centering
    \includegraphics[width=\linewidth]{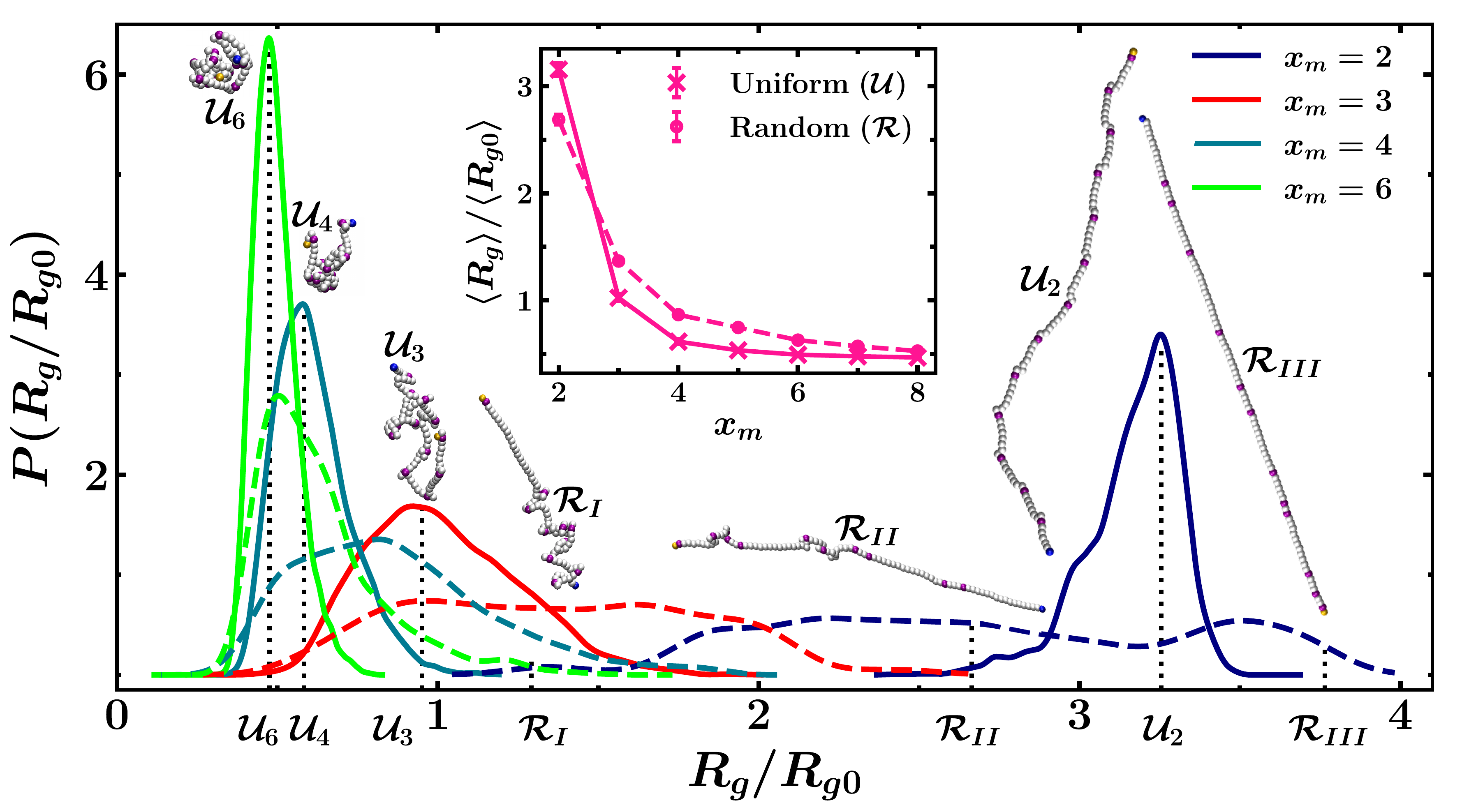}
     \caption{Probability distribution of the normalized radius of gyration for active polymers with uniformly~(solid lines) and randomly~(dashed lines) distributed motors with $\rho_m=0.125$, $N=100$ and Pe$=165$. The position of first motor is fixed at different positions~$x_m$ (see Movie3 in SM).  The inset plot shows the variation of $\langle R_g \rangle/\langle R_{g0}\rangle$ with~$x_m$. The snapshots represent polymer typical conformations, with radius of gyration indicated by the dotted lines. $\mathcal{U}$ and $\mathcal{R}$ denote respectively uniform and random configurations. Various values of the first motor position, $x_m$, are displayed for $\mathcal{U}_m$ configurations while all $\mathcal{R}$ have $x_m=2$ (see Movie4 and Movie5 in SM).
     }
     \label{fig:comparison}
\end{figure}
While the motor density has shown to critically influence the properties of polar polymers, how these motors are distributed along the chain is also crucial.
For a fixed motor density, $\rho_m=0.125$ at high activity value $\text{Pe}=165$, we compare the homogeneous distribution of motors considered until now, with a random distribution of motors, and sort the results according to the position of the first motor, denoted here as $x_m$.
For reference, we also perform simulations for the homogeneous case with varying location of the first motor.  
The inset of Fig.~\ref{fig:comparison} shows how the average radius of gyration drastically depends on $x_m$.  
When the first motor is at the head, $x_m=2$, the polymer is basically completely stretched,  while placing the first motor just one position behind, \textit{i. e.} $x_m=3$, shows a drastic difference that indicates the polymer is not anymore stretched but in a clearly much more coiled configuration. 
Increasing the number of passive monomers in front of the first motor, {\it i. e.} enlarging $x_m$, intensifies the trend of the polymer to coil into more compact structures.
Passive monomers in front of the first motor help the head to lose its directionality, which frequently randomizes the motion of the subsequent motors, eventually leading to coiled states. 

For the case with random distribution of motors, the differences among different configurations are large such that we evaluate the actual sizes by the radius of gyration probability distribution function, as shown in the main panel of Fig.~\ref{fig:comparison}. 
The plethora of possible random conformations clearly widens the $R_g$ distribution for the random configurations as compared to the uniform ones, which is especially drastic for $x_m=2$, when the first motor is placed at the head position. 
Snapshots of example configurations are depicted in Fig.~\ref{fig:comparison}, where $\mathcal{U}_m$ is denoting configurations with uniform distribution of motors at a given value of $x_m$. The difference between $\mathcal{U}_2$ (with $x_m=2$) and $\mathcal{U}_3$ (with $x_m=3$) illustrates the much more coiled structures when one single dangling monomer is placed in front of the first motor. The snapshots denoted as $\mathcal{R}$ are representative random configurations, all with $x_m=2$. $\mathcal{R}_{III}$ shows how an accumulation of motors close to the head leads to an overall stretched configuration, while $\mathcal{R}_{II}$ and $\mathcal{R}_I$ show that the accumulation at different places in the chain leads to coiled segments, resulting in more or less compact configurations. These effects persist for $x_m>2$, as indicated by the width of radius of gyration probability distributions, although the differences between random and homogeneous become less pronounced with increasing $x_m$, such that for $x_m=6$ only a reminiscent difference persists.  
Finally, note that although we have here focused on a particular set of parameters, we expect that qualitatively similar conclusions can be drawn for a wide range of values. 
Smaller values of $\text{Pe}$ are for example expected to show differences in stretching, although less obvious than the ones here discussed. Smaller values of $\rho_m$ are expected to show wider distributions of the radius of gyration, similar to larger values of $N$, which might also display a slightly different behavior with $x_m$.   

In conclusion, the density and location of the active sites play a major role in determining the conformational and dynamical behavior of polar active polymers.  
Compact coiled structures are to be found for polymers with a very large density of motors, and interestingly also for lower densities of motors but only for cases in which a number of passive dangling monomers are placed before the first motor. Otherwise, most structures are much more stretched than the equilibrium configurations. 
While the rotational characteristic times decrease with increasing local activity and number of motors, the self-propelled velocities also account for the total polymer extension, such that non-monotonic dependencies are found for large values of the local activity.  
Self-similarity is typically lost in the presence of polar activity, and this can change from coiled-head and stiff-tail for the cases with large density of motor, to the opposite stiff-head and coiled-tail for the cases with lower density of motors, independent of the local activity. 
These observations have numerous practical implications relevant in diverse fields, such as the design of synthetic phoretic active polymers, or the understanding of mechanisms related to molecular motors like RNA polymerase on bio-polymers. In particular, our results can explain the reason why biological relevant polymers like chromatin might change from a larger number of motors with lower activity to a  lower number of motors with high activity when the transcriptional activity needs to be increased. 

\begin{acknowledgments}
S. J. thanks a research fellowship from UGC India.
S. T. acknowledges SERB India for research grant number CRG/2022/003778.
The authors would like to acknowledge the HPC facility at IISER Bhopal for allocating the necessary computational resources.
The authors gratefully acknowledge the computing time granted by the JARA Vergabegremium and provided on the JARA Partition part of the supercomputer JURECA at Forschungszentrum J\"ulich~\cite{jureca21}.
\end{acknowledgments}


%

\end{document}